\documentclass[twocolumn,english,aps,prb,reprint,superscriptaddress,showpacs,longbibliography,showkeys]{revtex4-2}
\usepackage[dvipsnames]{xcolor}
\usepackage[colorlinks=True,linkcolor=blue,citecolor=magenta,linktocpage=True,pdfborder={0 0 0}]{hyperref}
\usepackage{graphicx}
\usepackage{adjustbox}
\usepackage{amsmath,amssymb}
\usepackage{comment}
\usepackage{stackengine}
\usepackage{multirow}
\usepackage{calc}
\usepackage{float}
\usepackage{subfigure}
\usepackage{tikz}
\usetikzlibrary{quantikz2}
\usepackage{afterpage}
\usepackage{booktabs}
\usepackage{bm}
\usepackage{siunitx} 
\usepackage{dcolumn}

\renewcommand{\bra}[1]{\left\langle#1\right|}
\renewcommand{\ket}[1]{\left|#1\right\rangle}
\newcommand{\ketbra}[1]{\ket{#1}\hspace{-0.25em}\bra{#1}}
\newcommand{\ketbrap}[2]{\ket{#1}\hspace{-0.25em}\bra{#2}}

\newcommand{\braketp}[2]{\left\langle#1\middle|#2\middle|#1\right\rangle}

\newcommand{\subfig}[2]{%
  \begin{minipage}[t]{\widthof{#2}}%
    \raggedright%
    {\tiny\textsf{\textbf{#1}}}\\[-0.1ex]%
    #2%
  \end{minipage}%
}

\begin{document}

\title{Multi-Qubit Parity Gates for Rydberg Atoms in Various Configurations}

\author{Javad Kazemi}
    \affiliation{Parity Quantum Computing Germany GmbH, Schauenburgerstraße 6, 20095 Hamburg, Germany}

\author{Michael Schuler}
    \affiliation{Parity Quantum Computing GmbH, Rennweg 1, Top 314, 6020 Innsbruck, Austria}

\author{Christian Ertler}
    \affiliation{Parity Quantum Computing Germany GmbH, Schauenburgerstraße 6, 20095 Hamburg, Germany}

\author{Wolfgang Lechner}
    \affiliation{Parity Quantum Computing Germany GmbH, Schauenburgerstraße 6, 20095 Hamburg, Germany}
    \affiliation{Parity Quantum Computing GmbH, Rennweg 1, Top 314, 6020 Innsbruck, Austria}
    \affiliation{Parity Quantum Computing France SAS, 10 Avenue de Kl{\'e}ber, 75016 Paris, France}
    \affiliation{Institute for Theoretical Physics, University of Innsbruck, 6020 Innsbruck, Austria}

\date{\today}

\begin{abstract}
We present a native approach for realizing multi-qubit parity phase gates in neutral atom systems through global phase modulation of a Rydberg excitation laser. By shaping the temporal profile of the laser’s phase, we enable high fidelity, time efficient entangling operations between multiple qubits without requiring individual qubit addressing. To mitigate intrinsic noise sources including spontaneous decay and motional effects, we develop a noise-aware optimal control framework that reduces gate errors under the presence of noise while maintaining smooth pulse profiles suitable for experimental implementation. In addition to equidistant qubit arrangements, we explore the impact of non-equidistant atomic configurations, where interaction inhomogeneity becomes significant. In these cases, the flexibility of our control approach helps to compensate for such variations, supporting reliable gate performance across different spatial layouts. These results facilitate the practical implementation of complex, multi-qubit quantum operations in near-term neutral atom quantum processors.
\end{abstract}

\maketitle

\section{Introduction}
The ability to perform multi-qubit quantum gates directly, without decomposing them into sequences of primitive operations, holds significant promise for advancing quantum computing. Traditionally, quantum circuits are constructed from a universal set of single- and two-qubit gates, which are well understood and relatively easier to implement. However, decomposing higher-order operations into such primitive gates leads to longer circuit depths and increased susceptibility to noise, particularly in near-term quantum devices where coherence times are limited. Direct implementation of multi-qubit gates can circumvent these issues by enabling faster execution with fewer error-prone operations, ultimately enhancing fidelity and reducing overhead \cite{Basilewitsch2024, Jandura2024}. 

Several theoretical and experimental works have explored the implementation of multi-qubit gates in different physical platforms. In particular, neutral-atom arrays provide a promising architecture for this purpose, thanks to their flexible connectivity and strong, controllable interactions via Rydberg excitations. Experimental demonstrations of multi-qubit gates have been reported in systems using Rydberg atoms such as rubidium \cite{Levine2019, Evered2023} and strontium \cite{Cao2024}, showcasing the feasibility of scaling up to more complex entangling operations.

In this work, we focus on a specific class of multi-qubit gates, namely parity phase gates, which have various applications in quantum computation and simulation. Families of parity phase gates play a key role in the simulation of complex quantum systems \cite{Barkoutsos2018, Maskara2025}, such as those described by lattice gauge theories, where multi-body interactions naturally arise \cite{Irmejs2023, Kalinowski2023, Mildenberger2025}. Beyond quantum simulation, parity-based operations are central to the concept of parity quantum computing and its applications \cite{Lechner2015, Fellner2022,Fellner2022b, Dlaska2022,Lanthaler2023, Weidinger2024, Messinger2024}, where it introduces an alternative computational approach for efficiently encoding arbitrary interaction graphs on quantum devices with only local connectivity.

In this paper, we consider neutral atom quantum processors to implement parity phase gates. This versatile quantum computing platform, that utilizes neutral atoms that are individually trapped in optical tweezers, has seen substantial progress in recent years, including the demonstration of high-fidelity operations~\cite{Evered2023, Radnaev2024, Tsai2025, Peper2025}, the ability to coherently reconfigure atom positions~\cite{Bluvstein2022, Maskara2025}, demonstrations of scalability to large qubit arrays with thousands of qubits~\cite{Manetsch2024, Norcia2024, Tao2024}, and even early demonstrations of quantum error correction~\cite{Bluvstein2024, Rodriguez2024, Reichardt2024, Bedalov2024}.

The paper is structured as follows:
In Sec.~\ref{sec:setup} we introduce the setup of the considered neutral atom QPU and our optimal control framework to design high-fidelity gate operations with smooth pulse profiles.
In Sec.~\ref{sec:speed_limit} we then analyze the minimal gate times for multi-qubit parity phase gates on different atom arrangements and analyze the effect of pulse smoothing regularization. 
To tackle intrinsic noise sources that could spoil gate fidelities, we introduce a noise-aware optimal control framework in Sec.~\ref{sec:noise_aware}. This allows us to maintain high gate fidelities even under the presence of fundamental noise sources such as Rydberg decay, photon recoil and van-der-Waals force induced motional effects.
Section~\ref{sec:errors} is devoted to a detailed study of the influence of different types of noise on the gate fidelities.
We analyze the effect of intrinsic noise under different external parameters and propose strategies to improve gate fidelities [Sec.~\ref{sec:errors_intrinsic}], we show that our native implementations of parity phase gates show biased noise profiles under dissipation-induced errors in stark contrast to corresponding circuits of single- and two-qubit gates [Sec.~\ref{sec:errors_tomography}], and finally we analyze the influence of technical noise sources on the gate performance [Sec.~\ref{sec:errors_technical}].
We conclude our work in Sec.~\ref{sec:conclusion} and provide additional information in the Appendix.

\section{Setup \& Optimal Control}\label{sec:setup}
We define the \textit{parity phase gate} acting on $N$ qubits as the unitary operation
\begin{equation}
    \mathrm{Z}_N(\theta) = \exp\left(-i\theta\, \sigma_1^{z} \otimes \sigma_2^z \otimes \cdots \otimes \sigma_N^{z}\right),
\end{equation}
where $\theta$ is a tunable parameter that sets the magnitude of the phase rotation and $\sigma_j^z$ denotes the Pauli-Z matrix on qubit $j$. This gate belongs to the broader class of multi-qubit phase (or diagonal) gates, which apply relative phase shifts to computational basis states without altering their bit values. The parity phase gate, in particular, applies a phase of $e^{-i\theta}$ to basis states with \textit{even parity}, i.e., an even number of $1$'s, and a phase of $e^{+i\theta}$ to those with \textit{odd parity}. The entangling power of $\mathrm{Z}_N(\theta)$ reaches its maximum when $\theta = (2n + 1)\pi/4$, where $n$ is an integer. 
In contrast, when $\theta = n\pi/2$, the gate reduces to a product of local $Z$-rotations and does not produce entanglement. In this work, we focus on the case of maximal entanglement with $\theta = \pi/4$ as it represents the most challenging scenario for experimental implementation and control.
Note that the parity gate implements a pure $N$-body interaction, unlike the multi-qubit controlled-phase gate
$\mathrm{C}_{N-1}\mathrm{P}(\theta) = \exp\left[ -i\theta\otimes_{i=1}^N (\mathrm{I}_i - \sigma_i^{z}) \right]$,
whose generator includes a mixture  of up to $N$-body interactions.


We implement the parity phase gate using neutral atom quantum processing units (QPUs), a platform that has seen tremendous progress in recent years. This platform's ability to arrange atoms in almost arbitrary configurations, combined with the intrinsic van-der-Waals interaction mechanism, allows for the native implementation of multi-qubit gates.
Specifically, for the atomic species, we focus on the isotope $^{88}\text{Sr}$, which is known for its rich electronic structure that enables laser cooling and control with high precision \cite{Cooper2018, Heinz2020, Finkelstein2024, Unnikrishnan2024, Ammenwerth2024, Shaw2025} and long coherence time \cite{Norcia2019, Young2020}.
These features made it a cornerstone in optical atomic clock technologies \cite{Takamoto2005, Bloom2014, Madjarov2019, Aeppli2024}. Its well-known optical-frequency clock transition can also be harnessed for quantum computing \cite{Schine2022, Finkelstein2024, Zhang2024}. We encode our logical $\ket{0}$ state in the electronic ground state \(\ket{5s^2\ ^1\text{S}_0}\), while \(\ket{1}\) is encoded in the long-lived metastable state \(\ket{5s5p\ ^3\text{P}_0}\). 
To generate entanglement between qubits, which is essential for the implementation of (non-trivial) multi-qubit gates, we assume a resonant laser field coupling the $\ket{1}$ state to the highly excited Rydberg state \(\ket{r} \equiv \ket{5s61s\ ^3\text{S}_1}\), giving rise to strong van-der-Waals (VdW) interactions between atoms in those states~\cite{browaeys2020}. 
We assume that the laser field addresses all atoms globally and that its Rabi frequency $\Omega(t)$ and laser phase $\phi(t)$ can be controlled.
The system’s evolution is described by the Hamiltonian
\begin{equation}\label{eq:ryd_ham}
\begin{split}
    H(t) =&\frac{\hbar\Omega(t)}{2}\sum_{j=1}^{N} \left(e^{i\phi(t)}\ketbrap{r_j}{1_j} + \mathrm{h.c.}\right) \\&-\sum_{j<k}\frac{C_6}{R_{jk}^{6}}n_jn_k,
    \end{split}
\end{equation}
where \(n_j=\ketbra{r_j}\) is the projector onto the Rydberg state of atom $j$. The last term describes the interaction between atoms in Rydberg states through strong van der Waals (VdW) forces. The interaction potential depends on
the interatomic spacing \(R_{jk}=|R_j - R_k|\) between atoms at positions $R_j$ and $R_k$, which is typically on the order of a few micrometers, and \(C_6 < 0\) indicates repulsive interaction. For the chosen Rydberg state \(\ket{5s61s\ ^3\text{S}_1}\) in $^{88}\text{Sr}$, the van-der-Waals coefficient is estimated as \(C_6/h \approx -150\, \mathrm{GHz}\,\mu\mathrm{m}^6\) using the \textit{pairinteraction} software \cite{Weber2017}. Given realistic experimental parameters, we assume maximum Rabi frequencies up to \(\Omega_0/(2\pi) = 10\,\mathrm{MHz}\) and minimal distances between atoms of \(R_{\text{min}} = 3\,\mu\mathrm{m}\). This yields a strong interaction regime with \(V/\hbar\Omega_{\max} \approx 21\) where $V=|C_6/R_{\min}^6|$, allowing for efficient entangling gate operations through dynamic modulation of the control parameters \(\{\Omega(t), \phi(t)\}\) \cite{Pagano2022}.

An important property of the Rydberg Hamiltonian Eq.~\eqref{eq:ryd_ham} is its block-diagonal structure, which arises because the laser only drives transitions between \(\ket{1}\) and \(\ket{r}\), leaving \(\ket{0}\) unaffected. As a result, the Hamiltonian can be decomposed as
\begin{equation}
    H = \sum_{\mu\in\{0,1\}^N} H_\mu = \sum_{\mu\in\{0,1\}^N} P_\mu H P_\mu,
\end{equation}
where $P_\mu$ is a projector that maps the state onto the subspace where the atom $i$ is in $\ket{0}$ if $\mu_i=0$, and in the subspace spanned by ${\ket{1}, \ket{r}}$ if $\mu_i=1$.
This decomposition enables independent evolution of each block $H_\mu$, significantly accelerating numerical simulations for optimal control. Note that spatial symmetries of the atomic arrangement can make subsets of the $H_\mu$ operators identical, which leads to a further reduction of simulations resources. 
It is important to note that the all-zero state \(\ket{00\ldots0}\) remains unaffected under the dynamics. Therefore, it is convenient to redefine the parity gate by absorbing a global phase factor, i.e. \(\mathrm{Z}_N(\theta) \rightarrow e^{i\theta(-1)^N} \mathrm{Z}_N(\theta)\).

\begin{figure*}[ht!]
\centering
    \begin{minipage}[t]{0.3\textwidth}
        \vspace{0.2cm}
        \subfig{(a)}{
        \includegraphics[width=\linewidth]{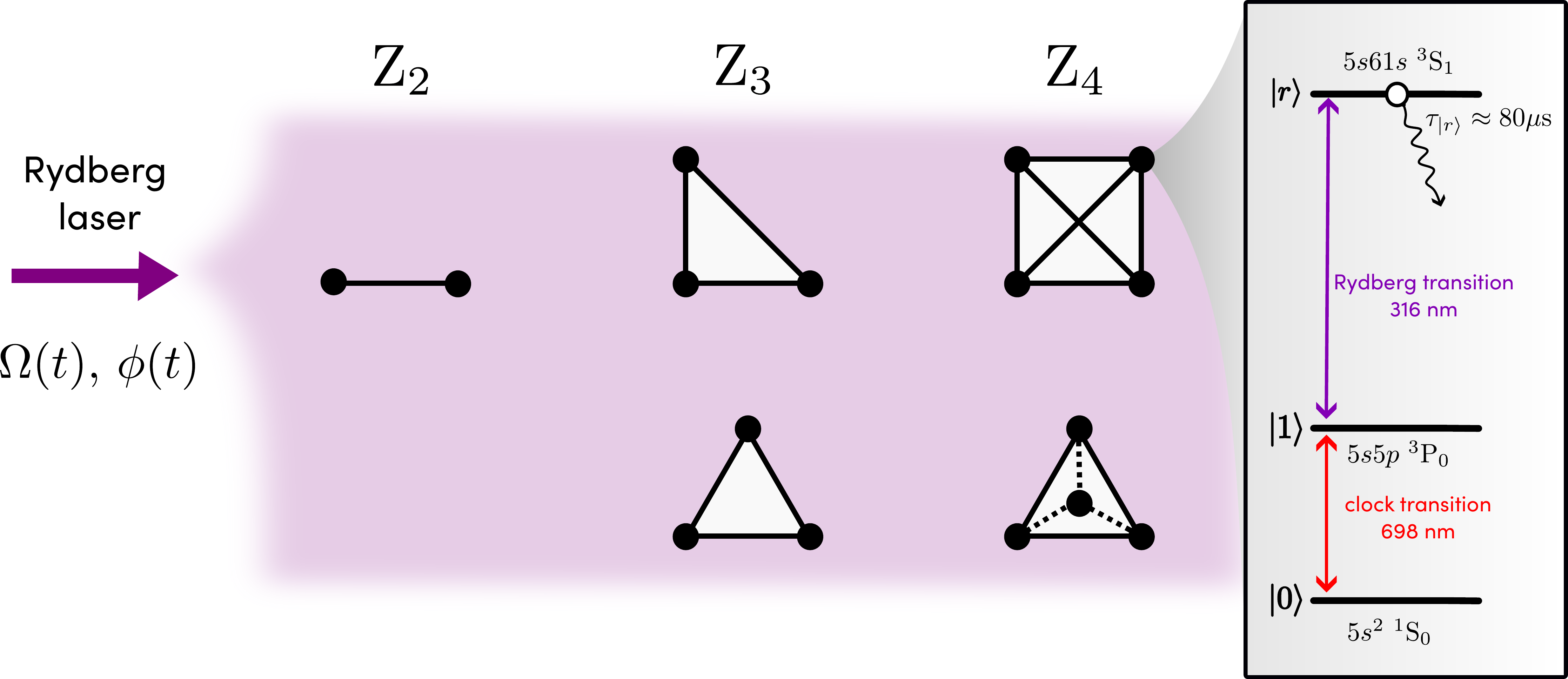}}\\
        \vspace{0.1cm}
        \subfig{(c)}{
        \includegraphics[width=\linewidth]{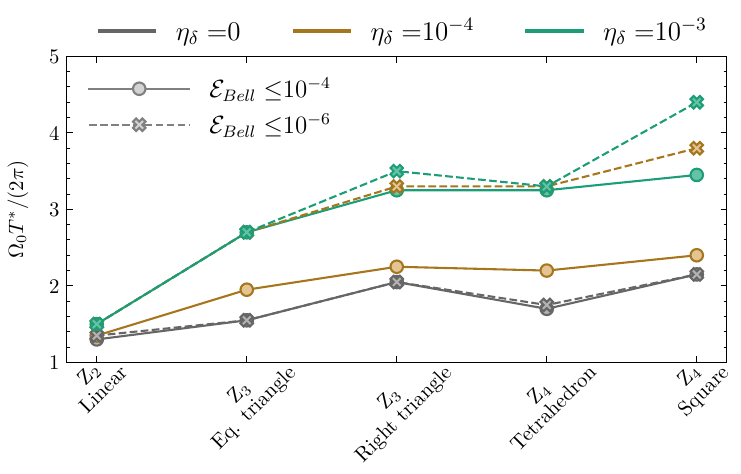}}
    \end{minipage}
    \hfill
    \begin{minipage}[t]{0.4\textwidth}
        \subfig{(b)}{
        \includegraphics[width=\linewidth]{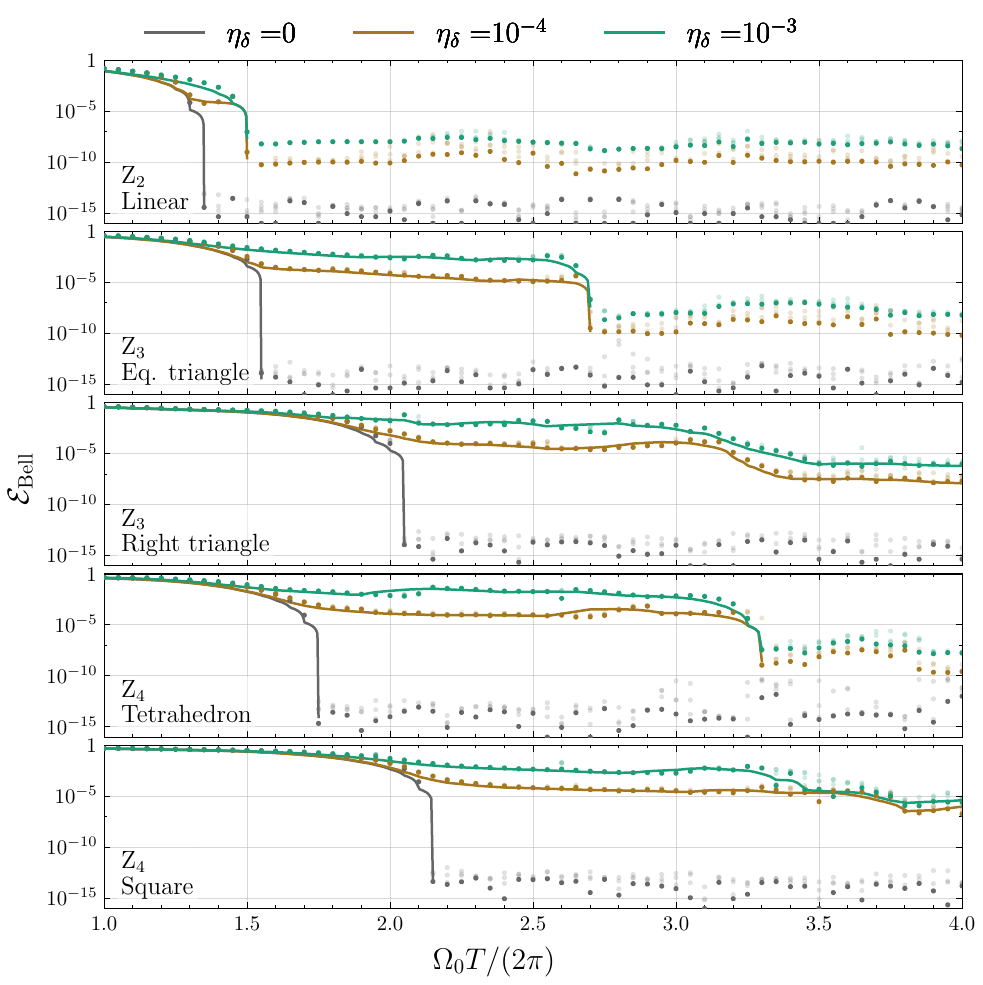}}
    \end{minipage}
    \hfill
    \begin{minipage}[t]{0.28\textwidth}
        \vspace{0.1cm}
        \subfig{(d)}{
        \includegraphics[width=\linewidth]{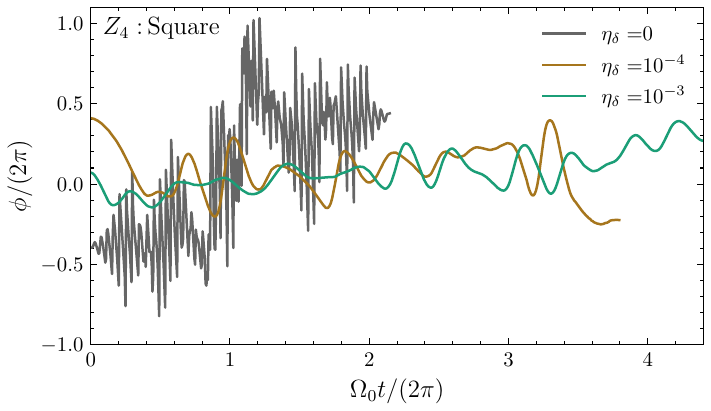}}\\
        \vspace{0.1cm}
        \subfig{(e)}{
        \includegraphics[width=\linewidth]{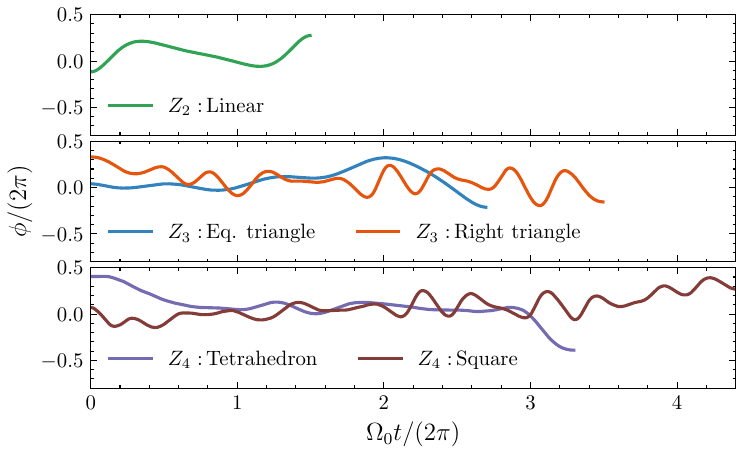}}
    \end{minipage}
    \caption{Optimal control results for multi-qubit phase gates $\mathrm{Z}_N(\theta)$ at their maximal entangling power $\theta=\pi/4$ implemented on different atomic configurations. 
    (a) Schematic of a global Rydberg laser driving neutral atoms arranged in different configurations. Qubits are encoded in the optical clock transition of the \({}^{88}\text{Sr}\) atom, with \(\ket{0} \equiv \ket{5s^2 \, ^1\text{S}_0}\) as the ground state and \(\ket{1} \equiv \ket{5s5p \, ^3\text{P}_0}\) as the metastable clock state. 
    Additionally, we consider excitation to the Rydberg state \(\ket{r} \equiv \ket{5s61s \, ^3\text{S}_1}\) to generate entanglement between qubits through the natural VdW force between atoms in Rydberg states.
    Pulse shaping is achieved through optimal control over the laser phase $\phi(t)$ and Rabi frequency $\Omega(t)$. 
    (b) Bell-state infidelity as a function of normalized gate duration \(\Omega_0 T/(2\pi)\) for various values of the regularization coefficient \(\eta_\delta\). Solutions without regularization (\(\eta_\delta = 0\), shown in gray) estimate the quantum speed limit, as indicated by an abrupt drop to near-perfect fidelity (\(\mathcal{E}_{\mathrm{Bell}} \leq 10^{-15}\)). Regularized solutions with \(\eta_\delta = 10^{-4}\) (brown) and \(10^{-3}\) (green) reveal the minimal gate durations achievable with smoother pulses. The brightly colored data points represent the best results out of 50 simulation runs for a given $T$, with the rest displayed faded-out for comparison. 
    (c) Minimal (normalized) gate duration \(\Omega_0 T^\ast/(2\pi)\) required to reach infidelity thresholds of \(10^{-6}\) (solid lines) and \(10^{-4}\) (dashed lines) across all studied configurations for different values of $\eta_{\delta}$. 
    (d) Phase profiles of time-optimal \(\mathrm{Z}_4\) pulses at the square configuration with \(\mathcal{E}_{\mathrm{Bell}} \leq 10^{-6}\), illustrating the effect of regularization strength \(\eta_\delta\). 
    (e) Phase profiles of time-optimal pulses for $\eta_\delta=10^{-3}$, showing that pulse shapes for equidistant configurations are simpler than for non-equidistant configurations.
    In all these results, only the laser phase $\phi(t)$ is optimized, while the Rabi frequency \(\Omega(t) = \Omega_0\) is kept constant.
    }
    \label{fig:qsl}
\end{figure*}

Our primary objective is to design control pulses $\{\Omega(t), \phi(t)\}$ that steer the system's evolution operator \( U(t) \) to closely replicate a target gate operation \( U_{\mathrm{target}} = \mathrm{Z}_N(\theta) \) after a duration $t=T$. A widely used figure of merit (FOM) for evaluating the performance of such gates is the Bell state fidelity, defined as
\begin{equation}\label{eq:bell_fidelity}
\mathcal{F}_{\mathrm{Bell}}(U) = \frac{1}{4^N}\left| \mathrm{Tr}\left[U_{\mathrm{target}}^\dagger U(T)\right] \right|^2.
\end{equation}
This measure captures how accurately the gate transforms the uniform superposition state 
\(
\ket{\psi_{\mathrm{init}}} = \frac{1}{\sqrt{2^N}} \sum_{\mu \in \{0,1\}^N} \ket{\mu}
\)
into the desired output state \( U_{\mathrm{target}} \ket{\psi_{\mathrm{init}}} \). For instance, when targeting the CZ gate or its equivalent parity phase gate \( \mathrm{Z}_2(\pi/4) \), the resulting state is a Bell state, motivating the name of this fidelity measure. Further details and comparisons with average gate fidelity are provided in Appendix~\ref{app:fidelity}.

To optimize the control pulses, we utilize a gradient-based quantum optimal control method based on the GRAPE (Gradient Ascent Pulse Engineering) framework. The control parameters \( u(t) = \{\Omega(t), \phi(t)\} \), representing the Rabi frequency and phase of a global laser, are modeled using a piecewise-constant ansatz (PCA). Within this framework, the total evolution time \( T \) is divided into \( M \) discrete pieces \( t_j=[j\Delta t, (j+1)\Delta t] \), with each interval characterized by length of $\Delta t = T/M$ and constant control values. This discretization enables approximating the total unitary evolution as a product of short-time propagators:
\begin{equation}
U \approx \prod_{j=0}^{M-1} \exp(-i H(t_{M-1-j}) \Delta t),
\end{equation}
where \( H(t_j) \) is the Hamiltonian evaluated at time piece $t_j$.

Optimization proceeds by iteratively updating the control parameters to minimize a cost function that penalizes deviation from ideal fidelity. We compute gradients of this cost using \texttt{JAX}, a high-performance numerical computing library that supports automatic differentiation~\cite{Jax2018}, enabling efficient and scalable optimization even in high-dimensional parameter spaces. The cost function we minimize is given by
\begin{equation}\label{eq:noise_free_cost}
\mathcal{C} = \mathcal{E}_{\mathrm{Bell}} + \eta_\delta \sum_{u \in \{\Omega, \phi\}} \sum_{j} \left( \frac{u(t_{j+1}) - u(t_j)}{2} \right)^2,
\end{equation}
where \( \mathcal{E}_{\mathrm{Bell}} = 1 - \mathcal{F}_{\mathrm{Bell}} \) quantifies gate infidelity, and the second term is a \textit{smoothness regularizer}. The regularization penalizes sharp variations in the control pulses, effectively encouraging bandwidth-limited solutions. This is particularly important given the limitations of current optical modulation hardware. The strength of this regularization is governed by the tunable weight \( \eta_\delta \). 
Together, these components guide the optimization toward physically implementable control solutions.
While smooth, experimentally feasible pulses can also be obtained using other techniques, such as initializing the optimization with a smooth profile, our approach – controlling pulse smoothness through regularization in the cost function – offers several advantages. It allows for precise control over the degree of smoothing via a single parameter, $\eta_\delta$, it enforces smoothness throughout the optimization process, and enables optimization with random initial guesses, mitigating the risk of getting trapped in local minima.

An important theoretical bound in quantum control is the \textit{quantum speed limit} (QSL), which defines the minimum time required for a quantum system to evolve between two distinguishable states. In the context of gate synthesis, this translates to a fundamental lower bound on the total gate duration, \( T^\ast \), necessary to implement a desired unitary operation with high fidelity. Estimating the QSL through computing minimal gate duration provides insight into the limits of gate performance and the efficiency of control strategies. 

\section{Minimal gate time}\label{sec:speed_limit}
In our work, we first numerically estimate the QSL for multi-qubit parity phase gates \(\mathrm{Z}_N(\theta)\) with \(N\in\{2, 3, 4\}\), examining a range of spatial configurations of neutral-atom qubits. These include both equidistant and non-equidistant configurations, which strongly affect the interaction pattern between the atoms as a result of the distance-dependent VdW interaction. To represent equidistant configurations, we consider atoms positioned at the vertices of an equilateral triangle and a regular tetrahedron for \(\mathrm{Z}_3(\theta)\) and \(\mathrm{Z}_4(\theta)\), respectively.
To represent non-equidistant configurations we consider planar arrangements, i.e. an isosceles right triangle and a square [see Fig.\ref{fig:qsl}(a)].

Figure~\ref{fig:qsl}(b) presents a summary of our optimal control results for the parity phase gate for all considered atom configurations, where we considered global phase modulation $\phi(t)$ but a fixed Rabi frequency $\Omega(t) = \Omega_0$. 
Each data point represents the resulting fidelity $\epsilon_{\rm Bell}$ of a single optimal control run for a given duration $T$. 
Without pulse regularization, \(\eta_\delta = 0\), one can clearly identify a sharp transition to the high-fidelity regime (indicated with solid lines) at a minimum duration \(T^\ast\), marking the intrinsic QSL for the given setup. 
Figure~\ref{fig:qsl}(c) directly compares the QSL durations \(T^\ast\) for the parity phase gate for different numbers of atoms and configurations, utilizing two distinct threshold values. 
For the non-regularized case, \(\eta_\delta=0\), one can clearly see that $T^\ast$ increases with $N$. Interestingly, the non-equidistant layouts have show a rather strongly increased threshold than the equidistant layouts.
We attribute this effect to the higher required control capabilities for non-equidistant layouts, where interactions of different magnitudes need to be properly synchronized for high-fidelity gate operations.

We also study the effect of finite pulse regularization on the results by varying the regularization weight \(\eta_\delta\). 
When a finite \(\eta_\delta\) is used to penalize high-frequency features in the control pulses, the QSL shifts to longer times [c.f. Figs.~\ref{fig:qsl}(b),(c)]. This effect is more pronounced in non-equidistant configurations where fidelity increases gradually with duration, indicating a smoother transition.
Importantly, even when the control pulses are strongly regularized to be smooth, the optimized multi-qubit gates remain significantly faster than equivalent circuits built from standard two-qubit gate decompositions. This suggests the practical benefit of using native multi-qubit gates for quantum computing.

As an illustrative example for the importance of pulse smoothness regularization of GRAPE pulses for experiments, Fig.~\ref{fig:qsl}(d) displays optimized phase control pulses for the square configuration for different regularization weights $\eta_\delta$, achieving high-fidelity gate performance \((\mathcal{E}_{\mathrm{Bell}} \leq 10^{-6})\). 
These results clearly show the effect of the regularization term in promoting smooth, low-bandwidth pulse shapes by suppressing rapid oscillations, at the cost of increasing the pulse duration. 
Furthermore, we observe that, in general, non-equidistant atomic configurations require more pronounced oscillations in the control pulses compared to their equidistant counterparts [see Fig.~\ref{fig:qsl}(e)]. This increased complexity is needed to compensate for the spatially varying interaction strengths while using a global Rydberg drive.

\section{Noise-Aware gate optimization}\label{sec:noise_aware}
So far, we have examined the implementation of Parity phase gate under idealized noise-free conditions. In practice, however, various noise sources reduce the achievable gate fidelity. In this section, we analyze fundamental, intrinsic noise sources of the envisioned experimental setup and describe how they can be incorporated into our optimal control framework to develop a noise-aware optimization strategy.

A significant source of intrinsic noise arises from the finite lifetime of the Rydberg state, leading to spontaneous decay. This decay imposes a fundamental limitation on the fidelity of entangling gates implemented with Rydberg atoms \cite{Wesenberg2007, Poole2025}. Importantly, its contribution to the average gate and Bell-state infidelity is known to scale linearly with the average time spent in the Rydberg state during a pulse, see details in Appendix~\ref{app:fidelity}. In other words, the infidelity of a perfect optimal pulse due to Rydberg decay can be expressed as 
\begin{equation}\label{eq:eps_decay}
    \mathcal{E}_{\mathrm{decay}} = \gamma_d T_R,
\end{equation}
where $\gamma_d^{-1} = 80\,\mu\mathrm{s}$ is the estimated Rydberg state lifetime for the chosen Rydberg state, and $T_R= \int_0^{T} \bar{n}_{R}(t) dt$ quantifies the time integration of Rydberg population averaged over all computational basis states:
\begin{equation}
    \bar{n}_{R}(t) = \frac{1}{2^N}\sum_{i=1}^{N}\sum_{\mu\in\{0,1\}^N} \braketp{\mu}{n_i(t)} \, .
\end{equation}
Here $n_i(t) = U^\dagger(t) n_i U(t)$ denotes the time-evolved projector onto the Rydberg state of atom $i$. Minimizing $T_R$ during the pulse optimization is therefore essential to mitigate Rydberg decay-induced errors.

Another intrinsic source of decoherence comes from coupling between the internal states of the atoms and the motional states in the optical tweezers, which arises from two main mechanisms~\cite{Robicheaux2021}.
First, each Rydberg excitation or de-excitation with a photon causes a momentum recoil to the atom, leading to fidelity loss that grows with $\mathcal{E}_{\mathrm{recoil}}\propto T_R^2$ \cite{Robicheaux2021}. Second, when two nearby atoms are simultaneously excited to Rydberg states, they experience a mutual force due to the gradient of the VdW interaction. This interaction causes atomic displacements that result in unwanted phase accumulation, thereby reducing gate fidelity \cite{Ates2013, Keating2015}. The corresponding error scales as $\mathcal{E}_{\mathrm{force}} \propto T_{RR}^2$, where $T_{RR}=\int_0^{T} \bar{n}_{RR}(t) dt $ quantifies the time spent in the doubly excited Rydberg state with state-averaged population $\bar{n}_{RR}(t)$ given by
\begin{equation}
    \bar{n}_{RR}(t) = \frac{1}{2^N}\sum_{i<j} \sum_{\mu\in\{0,1\}^N} \braketp{\mu}{n_i(t)n_j(t)} \, .
\end{equation}
Note that \( T_R \) represents the time spent in states that contain one or more Rydberg excitation, while \( T_{RR} \) corresponds to the time spent in states with at least two Rydberg excitations. This implies that \( T_R \geq T_{RR} \), since all configurations contributing to \( T_{RR} \) are also included in \( T_R \). We will take advantage of this property in our following analysis.

Taking these noise sources into account leads to a modified form of the effective Rydberg Hamiltonian including motional degrees of freedom
\begin{equation}\label{eq:spin-phonon_ham}
    \begin{split}
    H_{\mathrm{noisy}}(t) = &\frac{\hbar\Omega(t)}{2}\sum_{j=1}^{N} \left[e^{i\left(\phi(t)+\mathbf{k}\cdot \tilde{\mathbf{R}}_j\right)}\ketbrap{r_j}{1_j} + \text{h.c.}\right] \\
        &+ \sum_{j=1}^{N}\left[ \frac{\mathbf{P}_{j}^2}{2m} + \frac{1}{2}m\omega_{\parallel}^2\tilde{R}_{{j}_{\parallel}}^2 + \frac{1}{2}m\omega_{\perp}^2\tilde{R}_{j_{\perp}}^2 \right] \\
        &- i \frac{\hbar\gamma_d}{2} \sum_{j=1}^{N} n_j - \sum_{j<k} \frac{C_6}{|\mathbf{R}_{j} - \mathbf{R}_{k}|^6} n_j n_k,
    \end{split}
\end{equation}
where $\mathbf{P}_j$ and $\mathbf{R}_j$ are the momentum and position operators for the $j^\text{th}$ atom with mass $m$, each confined in an optical tweezer trap with $\omega_\parallel$ and $\omega_\perp$ denoting the trap frequencies along and perpendicular to the direction of the driving laser, respectively; The position operator $\tilde{\mathbf{R}}_j$ represents the relative position of atom $j$ with respect to the center of its trap and $\mathbf{k} = (2\pi/\lambda,0)$ is the wave vector associated with the Rydberg transition wavelength $\lambda = 323\,\mathrm{nm}$. For simplicity, we assume isotropic trapping \(\omega_\parallel=\omega_\perp\), with a frequency of $\omega/(2\pi) = 100\,\mathrm{kHz}$.

\begin{figure}[!t]
  \centering
  {\includegraphics[width=0.8\linewidth]{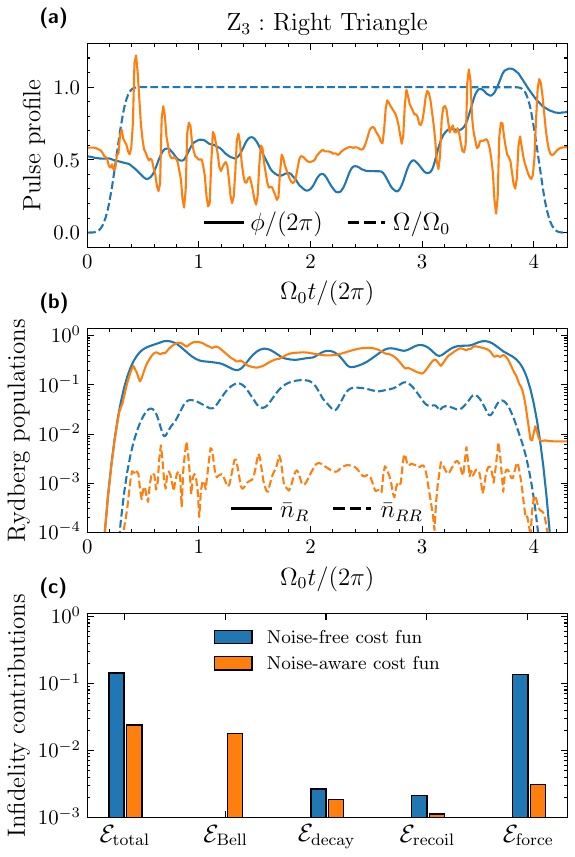}}%
\caption{
(a) Optimized pulses for the $Z_3$ gate in the right triangle atomic configuration, obtained using the noise-free cost function, Eq.~\eqref{eq:noise_free_cost}, (shown in blue) and the noise-aware cost function, Eq.~\eqref{eq:noise_aware_cost} (shown in orange).
(b) Time-resolved populations $\bar{n}_R$ and $\bar{n}_{RR}$ during gate operation. Using the noise-aware cost function (orange), which includes a regularization of the Rydberg times $T_R$ and $T_{RR}$, clearly reduces the population of Rydberg states (especially $n=2$) throughout the gate.
(c) Error budget showing the trade-off between contributions from intrinsic noise sources ($\varepsilon_{\rm decay}$, $\varepsilon_{\rm recoil}$, $\varepsilon_{\rm force}$) and the ideal infidelity $\varepsilon_{\rm Bell}$. The total infidelity $\varepsilon_{\rm total}$ is approximately equal to the sum of all these contributions (see main text for details).
}
\label{fig:different_cost_funs}%
\end{figure}

Note that Rydberg decay can be simulated efficiently using the non-Hermitian term $-i \hbar \gamma_d n_j / 2$ in the Hamiltonian, Eq.~\eqref{eq:spin-phonon_ham}, circumventing the need for full Lindblad master equation simulations. 
However, incorporating motional effects necessitates to add motional modes to the Hilbert space. To manage the basically infinite Hilbert space dimension, we only consider one-dimensional trapping, either parallel or orthogonal to the direction of the driving laser, and we restrict the simulations to a finite number of motional modes. Also, we employ a Krylov subspace solver to compute the time evolution of $H_{\mathrm{noisy}}(t)$~\cite{Nandy2025}.

We have shown above, that the considered noise sources impact Bell-state fidelity directly through $T_R$ and $T_{RR}$. Therefore, we include those quantities into the optimization cost function to search for noise-aware control pulses.
This strategy allows us to account for noise without requiring full Hamiltonian simulations at each optimization step. Specifically,  we define a \textit{noise-aware} cost function as
\begin{equation}\label{eq:noise_aware_cost}
    \mathcal{C}_{\mathrm{noisy}} = \mathcal{C} + \eta_R T_R + \eta_{RR} T_{RR},
\end{equation}
where $\eta_R$ and $\eta_{RR}$ are regularization weights that penalize the time spent in the singly and doubly excited Rydberg states, respectively.
A finite value of $T_R$ is required to mediate interactions between atoms and thus essential to create entangling gates. Finite $T_{RR}$, on the other side, is not essential and it can vanish, for example under perfect (infinite) blockade conditions. Therefore, it is appropriate to set the regularization weights such that $\eta_{RR} \gg \eta_{R}$.
For the rest of this work, we therefore use the regularization values \(\{\eta_\delta, \eta_R, \eta_{RR}\} = \{10^{-3}, 10^{-2}, 10^{2}\}\). The modified cost function, Eq.~\eqref{eq:noise_aware_cost}, ensures that the optimized control pulses remain smooth and achieve high-fidelity gate operations despite the presence of intrinsic noise, unlike the \textit{noise-free} cost function in Eq.~\eqref{eq:noise_free_cost}, which does not penalize Rydberg population.

Furthermore, to identify optimal control pulses suited for experimental implementation, one needs to consider a smooth turning-on and turning-off of the Rabi frequency $\Omega(t)$.
To achieve this we divide $\Omega(t)$ into three parts: First, we smoothly ramp up the Rabi frequency from $\Omega=0$ to $\Omega=\Omega_{\max}$ following the function \(\Omega_{\max} \left[1- e^{-\kappa (t/\tau_{\mathrm{ramp}})^4}\right]\), applied over a duration of $\tau_{\mathrm{ramp}} = \pi / \Omega_{\max}$ with the steepness parameter $\kappa=10$. After that, the Rabi frequency is kept constant at the value $\Omega_{\max}$ for a duration of $T-2\tau_{\rm ramp}$, before it is finally symmetrically ramped down to zero by the inverse of the ramp-up pulse profile [see Fig.~\ref{fig:different_cost_funs}(a) for an example].
These smooth ramps lead to an increase in the minimal gate duration by at most $2\pi / \Omega_{\max}$. 

To analyse the quality of noise-aware optimal control pulses obtained with the above mentioned procedure, we evaluate the gate performance under realistic noise conditions. In particular,
we initialize the quantum system in the motional ground state of the optical tweezers, i.e. zero temperature, with the internal state prepared in the uniform superposition state \(\ket{\psi_{\mathrm{init}}}\). 
The time evolution is then simulated by numerically solving the Schrödinger equation using the noisy Hamiltonian \(H_{\mathrm{noisy}}(t)\), Eq.~\eqref{eq:spin-phonon_ham}, with a Krylov subspace method. The simulations include up to 12 motional states per atom and retain terms up to $10^{\mathrm{th}}$ order in the Taylor expansion of the VdW potential \(1/R^6\).
The gate infidelity under these conditions is computed as \(\mathcal{E}_{\mathrm{total}} = 1 - \mathcal{F}_{\mathrm{Bell}}(U_{\mathrm{noisy}})\), providing a realistic measure of performance that accounts for intrinsic noise. Notably, in the parameter regime explored here, the contributions from different noise mechanisms are approximately additive
\begin{equation}
\mathcal{E}_{\mathrm{total}} \approx \mathcal{E}_{\mathrm{Bell}} + \mathcal{E}_{\mathrm{decay}} + \mathcal{E}_{\mathrm{recoil}} + \mathcal{E}_{\mathrm{force}}.
\end{equation}
This additive behavior supports our effective one-dimensional treatment of the trapping potential.

With the framework established, we now examine the impact of noise-aware optimization by analyzing optimal control pulses for the \(\mathrm{Z}_3\) gate in the right-triangle atomic configuration. Figure~\ref{fig:different_cost_funs}(a) shows two pulses of equal duration, obtained by minimizing the noise-free and noise-aware cost functions, respectively. 
Figure~\ref{fig:different_cost_funs}(b) shows a reduction of the Rydberg populations $\bar{n}_R(t)$ and $\bar{n}_{RR}(t)$ for the noise-aware optimization. Integrated over the full pulse, this leads to a reduction of $T_R$ by about $27\%$ and of $T_{RR}$ by almost $97\%$ compared to the noise-free result.
These reductions translate into a decrease in all three dominant noise contributions (Rydbderg decay, Photon recoil, VdW force), while, contrary, the pure gate infidelity $\epsilon_{\rm Bell}$ is increased. Nevertheless, the overall gate infidelity $\epsilon_{\rm total}$ is decreased by almost an order of magnitude, owing to the strong reduction of the noise contributions. This is shown in Fig.~\ref{fig:different_cost_funs}(c). 
In equidistant configurations, such improvements are achieved without a significant trade-off. However, in non-equidistant configurations [as shown in Fig.~\ref{fig:different_cost_funs}(c)], a more pronounced compromise arises between \(\mathcal{E}_{\mathrm{Bell}}\) and \(\mathcal{E}_{\mathrm{force}}\). This effect originates from the inhomogeneous phase accumulation caused by non-uniform interaction strengths under global driving. While the noise-free optimization tends to mitigate this by allowing increased time in doubly excited states, the noise-aware approach manages the trade-off by balancing contributions from different error sources [c.f. Fig.~\ref{fig:different_cost_funs}(c)].

\begin{figure}[!t]
    \centering{\includegraphics[width=0.48\linewidth]{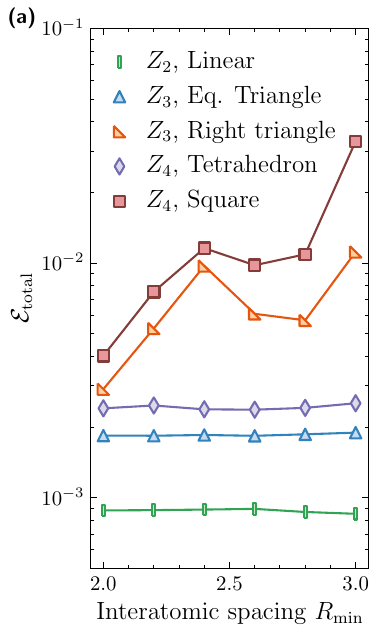}}
  {\includegraphics[width=0.48\linewidth]{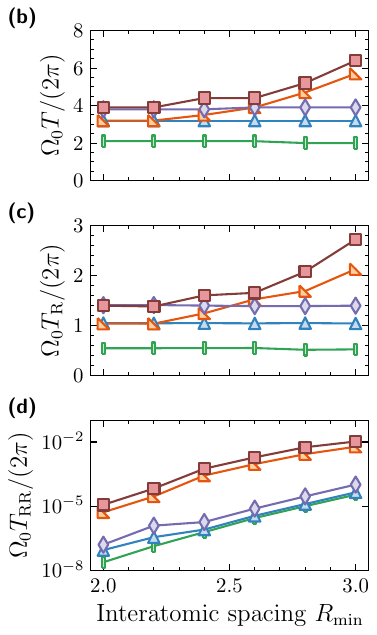}}
\caption{
(a) Total gate infidelity obtained from the noise-aware optimal control framework (see main text for details), along with the time budget components: (b) gate duration \(T\), (c) Rydberg time \(T_R\), and (d) Rydberg time \(T_{RR}\), all plotted as functions of interatomic spacing $R_{\min}$. For each data point, we report the lowest infidelity obtained from 20 optimization runs, subject to the constraint \(\Omega_0 T / (2\pi) \leq 7.0\).
}
\label{fig:tot_infidelity}
\end{figure}

Next, we investigate the influence of interatomic spacing \( R_{\min} \) on gate performance. Figure~\ref{fig:tot_infidelity}(a) shows the behavior of \(\mathcal{E}_{\mathrm{total}}\) as a function of interatomic spacing, varied down to \(2\,\mu\mathrm{m}\). Such closely spaced arrays are experimentally accessible using optical lattices \cite{Cao2024} or optical tweezers with high numerical aperture lenses \cite{Chew2022, Evered2023}. The results suggest that for equidistant configurations, gate fidelity remains relatively stable across the range \(R_{\min} \in [2, 3]\,\mu\mathrm{m}\), while the performance of non-equidistant configurations (right triangle, square) improves for smaller separations. This behavior is consistent with the total time budget for the gate, including the gate duration \(T\) and the time spent in Rydberg states [see Fig.~\ref{fig:tot_infidelity}(b-d)]. 
Both \(T\) and \(T_R\) increase generally with qubit number $N$, but show similar values for equidistant and non-equidistant atom configurations, especially for small interatomic spacings [see Fig.~\ref{fig:tot_infidelity}(b,c)]. 
The non-equidistant configurations, however, show larger slopes of $T$ and $T_R$ when $R_{\min}$ is increased, leading to somewhat longer pulse durations at $R_{\min}=3 \mu\rm{m}$, compared to the equidistant configurations. 
In stark contrast, \(T_{RR}\) shows a substantial difference between equidistant and non-equidistant configurations, the latter having several orders of magnitude larger values throughout all interatomic spacings, as shown in Fig.~\ref{fig:tot_infidelity}(d). 
Overall, enhancing interaction strength, either by accessing higher Rydberg states to increase \(C_6\), or by reducing interatomic separation, can mitigate VdW-induced errors and reduce the fidelity gap between equidistant and non-equidistant configurations.

\begin{figure}[!t]
\centering\includegraphics[width=0.9\linewidth]{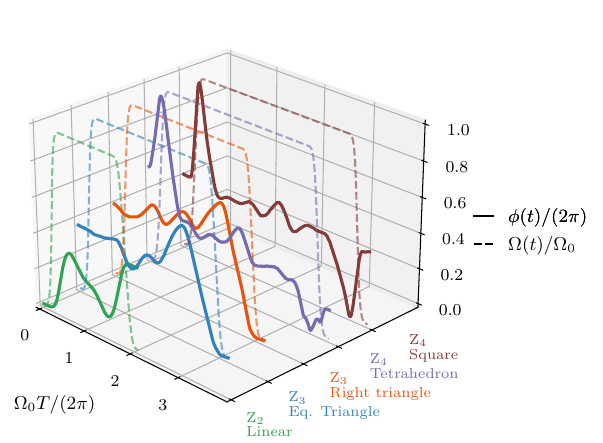}
\caption{
Optimized pulse profiles implementing multi-qubit parity gates for various atomic configurations, obtained through laser phase modulation (solid lines) and smooth ramping of the Rabi frequency \(\Omega_0\) (dashed lines). The pulses correspond to the results shown in Fig.~\ref{fig:tot_infidelity} at \(R_{\min} = 2\,\mathrm{\mu m}\).
}
\label{fig:opt_pulses}
\end{figure}

Figure~\ref{fig:opt_pulses} depicts representative optimal pulses for various configurations, based on our earlier findings at an interatomic distance of \(R_{\min} = 2\,\mu\mathrm{m}\). The pulse for the \(\mathrm{Z}_2\) gate recovers the near-sinusoidal shape characteristic of time-optimal \(\mathrm{CZ}\) gates \cite{Jandura2023, Evered2023}. As the qubit number \(N\) increases, the pulse profiles become moderately more complex, yet retain a smooth structure compatible with available experimental capabilities, due to the regularizations used in our approach.

\section{Error analysis}\label{sec:errors}
\sisetup{table-align-text-post=false}
\afterpage{
\begin{table*}[!t]
\centering
\caption{Error budget and time budget for $Z_N$ gates on different configurations with interatomic spacing of $R_{\min} = 2 \, \mathrm{\mu m}$.}\label{tab:error_budget}
\small
\begin{tabular}{@{}l@{\hskip -10pt}S@{\hskip -10pt}S@{\hskip -10pt}S@{\hskip -10pt}S@{\hskip -10pt}S@{}}
\toprule
\midrule
  & \hspace{10mm}{$Z_2: \mathrm{Linear}$} & \hspace{10mm}{$Z_3: \mathrm{Eq.\, Triangle}$} & \hspace{10mm}{$Z_3: \mathrm{Right\, Triangle}$} & \hspace{10mm}{$Z_4: \mathrm{Tetrahedron}$} & \hspace{10mm}{$Z_4: \mathrm{Square}$} \\
\midrule
Noiseless infidelity $\mathcal{E}_{\rm Bell}$  & 2.09e-8 & 2.45e-8 & 1.10e-3 & 2.13e-7 & 1.63e-3 \\
Rydberg decay $\mathcal{E}_{\rm decay}$          & 6.82e-4 & 1.31e-3 & 1.29e-3 & 1.76e-3 & 1.74e-3 \\
Photon recoil $\mathcal{E}_{\rm recoil}$         & 2.03e-4 & 5.28e-4 & 4.85e-4 & 6.44e-4 & 6.16e-4 \\
VdW force  $\mathcal{E}_{\rm force}$             & 1.76e-7 & 2.17e-6 & 3.40e-5 & 1.53e-5 & 4.77e-5 \\
Full simulation $\mathcal{E}_{\rm total}$        & 8.83e-4 & 1.84e-3 & 2.91e-3 & 2.42e-3 & 4.03e-3 \\
\midrule
$\Omega_0T/(2\pi)$         & 2.1 & 3.2 & 3.2 & 3.8 & 3.9 \\
$\Omega_0T_R/(2\pi)$         & 5.46e-1 & 1.05 & 1.04 & 1.41 & 1.40 \\
$\Omega_0T_{RR}/(2\pi)$         & 2.44e-8 & 9.04e-8 & 5.26e-6 & 2.45e-7 & 1.17e-5 \\
\midrule
\bottomrule
\end{tabular}
\end{table*}
}
\subsection{Infidelities from intrinsic noise}\label{sec:errors_intrinsic}

\begin{figure}[!t]
  \centering
  \includegraphics[width=0.95\linewidth]{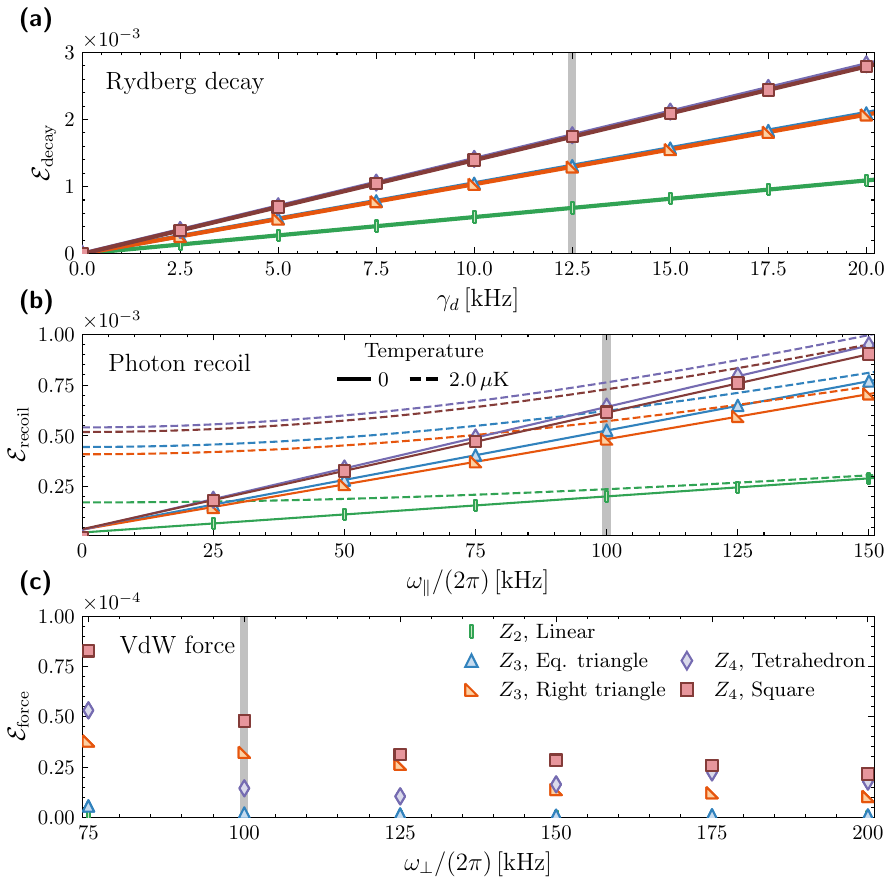}%
      \caption
    {%
      Infidelity contributions of intrinsic noise sources
      for different atomic configurations. 
      (a) Infidelity caused by Rydberg decay over a range of decay rates \(\gamma_d\). Data points are obtained numerically, while solid lines represent the analytical approximation \(\gamma_d T_R\). 
      (b) Infidelity due to photon recoil for different trapping frequencies $\omega_{\parallel}$ along the direction of the driving laser. 
      Solid lines correspond to fits of Eq.~\eqref{eq:recoil_analytical} at zero temperature (\(1/\beta=0\)) with fitting parameter $\alpha$. 
      Dashed lines represent extrapolations of the fitted results to finite temperature $1/\beta = 2 \, \mu\text{K}$.
      (c) Infidelity from the VdW force for different trapping frequencies $\omega_{\perp}$ orthogonal to the direction of the driving laser.
      In all plots, gray vertical lines indicate the values of the decay rate and trapping frequencies considered in the rest of the paper.
    }
\label{fig:noise_sources}%
\end{figure}

Focusing on the set of pulses shown in Fig.~\ref{fig:opt_pulses}, we now investigate the contributions of experimental noise sources in more detail.
We perform numerical simulations to assess the dependence of the gate error on decay rates and trap frequencies. Figure~\ref{fig:noise_sources}(a) confirms the expected linear scaling of the decay-induced error with the decay rate $\gamma_d$ [c.f. Eq.~\eqref{eq:eps_decay}]. 

The numerically obtained infidelity from photon recoil at zero temperature, $\mathcal{E}_{\rm recoil}$, is shown in Fig.~\ref{fig:noise_sources}(b). The data aligns well with the analytical estimate  
\begin{equation}
\mathcal{E}_\mathrm{recoil} = \alpha \frac{\hbar k^2}{2m}\omega_{\parallel}T_R^2 \coth\left(\frac{\hbar \beta \omega_{\parallel}}{2}\right)
\label{eq:recoil_analytical}
\end{equation}
for all considered atom arrangements, where \(\beta\) denotes the inverse temperature~\cite{Pagano2022}.
$\alpha$ is treated as a fitting parameter and obtained by fitting the numerically calculated results, individually for each atom arrangement, at zero temperature, $1/\beta=0$ [see solid lines].
Using Eq.~\eqref{eq:recoil_analytical} and the results for the fitting parameters one can then extrapolate the infidelities from photon recoil also for finite temperatures of the atoms in the traps.
The dashed lines in Fig.~\ref{fig:noise_sources}(b) show such an extrapolation for a temperature of $2\,\mu{\rm K}$. We see that the infidelity from photon recoil increases slightly with temperature for the here considered target value of the the trapping frequency $\omega_\parallel$ [see gray vertical line], but the effect becomes stronger for smaller $\omega_\parallel$.

Figure~\ref{fig:noise_sources}(c), finally, shows the contribution of the infidelity due to VdW-induced forces between the atoms, $\mathcal{E}_{\rm force}$. In contrast to $\mathcal{E}_{\rm recoil}$ the infidelity contribution mostly decreases with increasing trap frequency $\omega_\perp$, although we observe non-monotonic behavior for some of the atom configurations.
We were unable to obtain satisfactory results when attempting to fit our data to the analytical approximation \(
\mathcal{E}_\mathrm{force} = \alpha^\prime \frac{1}{2m\hbar \omega_{\perp}}\left[\frac{C_6 T_{RR}}{R^7}\right]^2 \coth\left(\frac{\hbar \beta \omega_{\perp}}{2}\right)
\)~\cite{Pagano2022} at zero temperature with fitting parameter $\alpha^\prime$. This indicates that higher-order corrections might have to be included in the analytical formulation for multi-qubit gates.

A comprehensive numerical error and time budget is summarized in Table~\ref{tab:error_budget} for this set of pulses. The data show that fidelities that approach \(F = 0.999\) are achievable under realistic conditions for native multi-body parity phase gates. Motional errors ($\mathcal{E}_{\rm recoil}$ and $\mathcal{E}_{\rm force}$) are well suppressed, leaving Rydberg decay as the primary intrinsic noise source.
The time budget shows that the gate durations $T$ correspond to performing approximately $N$ Rabi cycles, demonstrating a highly efficient native gate implementation for many-body gates.
Furthermore, noise-aware optimization keeps $T_R$ minimal, with values ranging from \(T_R/T = 0.26\) for the \(\mathrm{Z}_2\) gate to \(T_R/T = 0.36\) for the \(\mathrm{Z}_4\) gate, while at the same time significantly reducing \(T_{RR}\). Additional strategies for improving the fidelity by increasing the Rabi frequency are discussed in Appendix~\ref{app:rabi}.

\subsection{Error tomography for dissipation-induced errors}\label{sec:errors_tomography}
After the detailed analysis of infidelity contributions from different noise sources, we now continue with a tomographic investigation of the noise profiles for the parity phase gates in terms of Pauli error channels.
In particular, we focus on dissipation-induced errors using the error process matrix formalism~\cite{Korotkov2013}.

\begin{figure}[!]
    \subfig{\hspace{-3mm}(a) \hspace{2.5mm} native}{
        \begin{adjustbox}{height=0.65cm}
        \input{circuits/Z4_native}
        \end{adjustbox}
        } 
    \subfig{\hspace{6mm} V decomposition}{
        \begin{adjustbox}{height=0.65cm}
        \input{circuits/Z4_V_decomposition}
        \end{adjustbox}
        }
    \subfig{\hspace{4mm} X decomposition}{
        \begin{adjustbox}{height=0.65cm}
        \input{circuits/Z4_X_decomposition}
        \end{adjustbox}
        }
    \subfig{\hspace{-1mm} ZZ decomposition}{
        \begin{adjustbox}{height=0.65cm}
        \input{circuits/Z4_ZZ_decomposition}
        \end{adjustbox}
        } \\
    \hspace{22mm}
    \subfig{}{
    \includegraphics[width=0.98\linewidth]{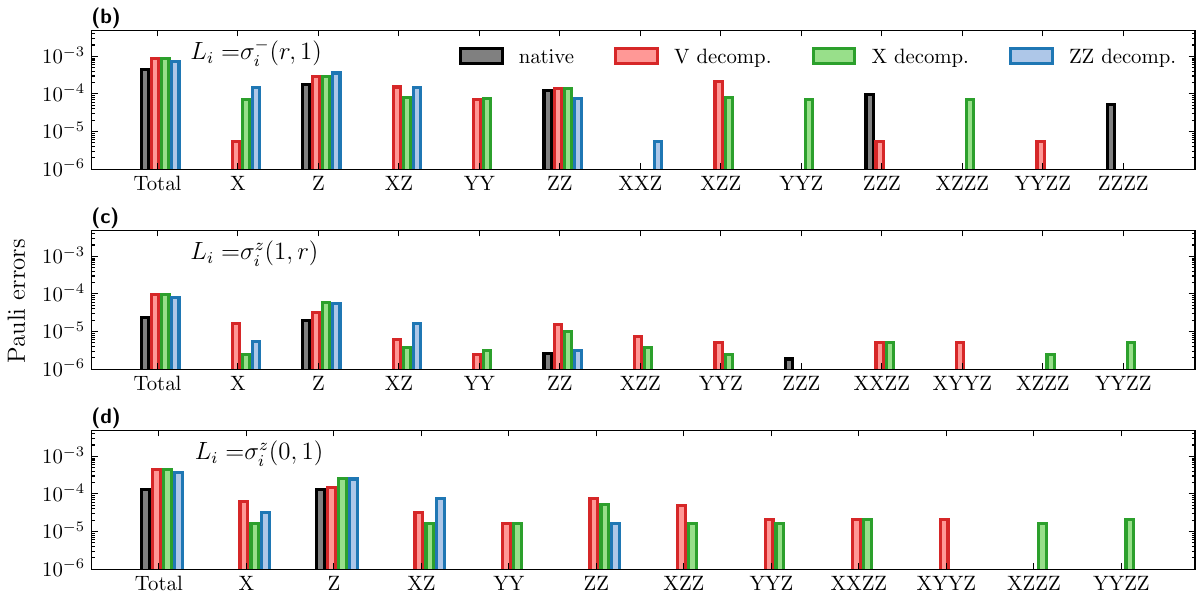}
    }

    \caption{
Dominant Pauli error contributions for different circuit implementations of the parity gate $Z_4$ (a), under various dissipation mechanisms. Dissipation is modeled using different local jump operators \(L_i\) (b-d) acting independently on each atom \(i\), with a uniform rate \(\gamma\). Only Pauli errors with magnitudes larger than \(10^{-6}\) are shown, with each dominant contribution representing the sum of identical Pauli errors across all atoms. 
(b) Pauli errors due to Rydberg state decay with a rate \(\gamma_d = 12.5\,\mathrm{kHz}\). 
(c, d) Pauli errors due to dephasing processes in the \(\{\ket{1}, \ket{r}\}\) and \(\{\ket{0}, \ket{1}\}\) subspaces, respectively, each with \(\gamma_d = 0.1\,\mathrm{kHz}\). The total error indicates the sum of all Pauli error contributions.
}
\label{fig:pauli_errors}
\end{figure}

For weakly dissipative dynamics governed by jump operators $L_k$ with corresponding decay rates $\gamma_k$, the diagonal components $p_m$ of the error process matrix are given by
\begin{equation}\label{eq:pauli_error}
    p_m = \frac{1}{4^{N}}\sum_k \gamma_k \int_0^T dt\, \left| \mathrm{Tr}\left[ U^\dagger(t) L_k U(t)E_m^\dagger\right]\right|^2 + \mathcal{O}( \gamma_k^2T^2),
\end{equation}
where \(\{E_m\}\) denotes the Pauli operator basis [see Appendix~\ref{app:pauli_errors} for details].
These diagonal components correspond to the probability that the Pauli error $E_m$ appears during the gate execution.
The approximation, Eq.~\eqref{eq:pauli_error}, is valid in the limit of weak dissipation, \(\gamma_k T \ll 1\), which is appropriate for the high-fidelity gates considered in this work. While a dissipative operation, in general, also gives rise to off-diagonal elements in the error process matrix, we restrict our analysis to the diagonal components to examine the types of Pauli errors that dominate.
To this end, we compare the native implementation of the \(\mathrm{Z}_4\) parity gate with different decompositions of this gate into a conventional gate set with only one- and two-body gates, as shown in Fig.~\ref{fig:pauli_errors}(a).
In the decompositions, we assume access to arbitrary, noise-free single-qubit rotations and use the \(\mathrm{Z}_2\) gate presented earlier as entangling gate, which is subject to noise. 

We begin by analyzing errors due to spontaneous decay from the Rydberg state which, as we have shown above, is the dominant source of noise in our setup. This is now modeled using jump operators \(L_i = \sigma_{i}^{-}(r,1) = \ketbrap{1_i}{r_i}\) with a decay rate \(\gamma_d = 1/80\,\mu\mathrm{s}^{-1}\). 
Figure~\ref{fig:pauli_errors}(b) shows that the total error rate (the sum of all Pauli error probabilities) for the native implementation of the $\mathrm{Z}_4$ is lower than for all decompositions. 
Even more important, the native implementation exhibits a pronounced bias towards \(Z\)-type Pauli errors, whereas the decomposed gate implementations lead to more diverse error distributions containing $X$ and $Y$-type Pauli errors.
Also note that inevitable Rydberg decay processes to other levels outside the computational subspace contribute only at higher orders in \(\gamma_d T\) and are therefore negligible within the current framework.

Next, we examine Pauli errors resulting from dephasing, modeled as separate channels in the \(\{\ket{1}, \ket{r}\}\) and \(\{\ket{0}, \ket{1}\}\) subspaces, with corresponding jump operators \(L_i = \sigma_i^{z}(1, r) = \ketbrap{1}{1} - \ketbrap{r}{r}\) and \(L_i = \sigma_i^{z}(0, 1) = \ketbrap{0}{0} - \ketbrap{1}{1}\), respectively. In both cases, we consider a dephasing rate of \(\gamma_i = 0.1\,\mathrm{kHz}\). Figures~\ref{fig:pauli_errors}(c, d) demonstrate that the native implementation, again, has lowest total error and yields only \(Z\)-type errors, whereas the decomposed circuits introduce additional \(X\)- and \(Y\)-type errors. These arise primarily from the use of single-qubit rotations interleaved with entangling two-qubit gates. 

In summary, the native \(\mathrm{Z}_4\) gate not only offers improved fidelity under noise but also displays a strongly biased Pauli error profile. Such a bias is advantageous in the context of quantum error correction, where tailored codes can exploit asymmetric error channels to enhance logical performance \cite{Sahay2023, Messinger2024, Tiurev2025}.

\subsection{Technical noise}\label{sec:errors_technical}

\begin{figure}[!t]
  \centering
\includegraphics[width=0.85\linewidth]{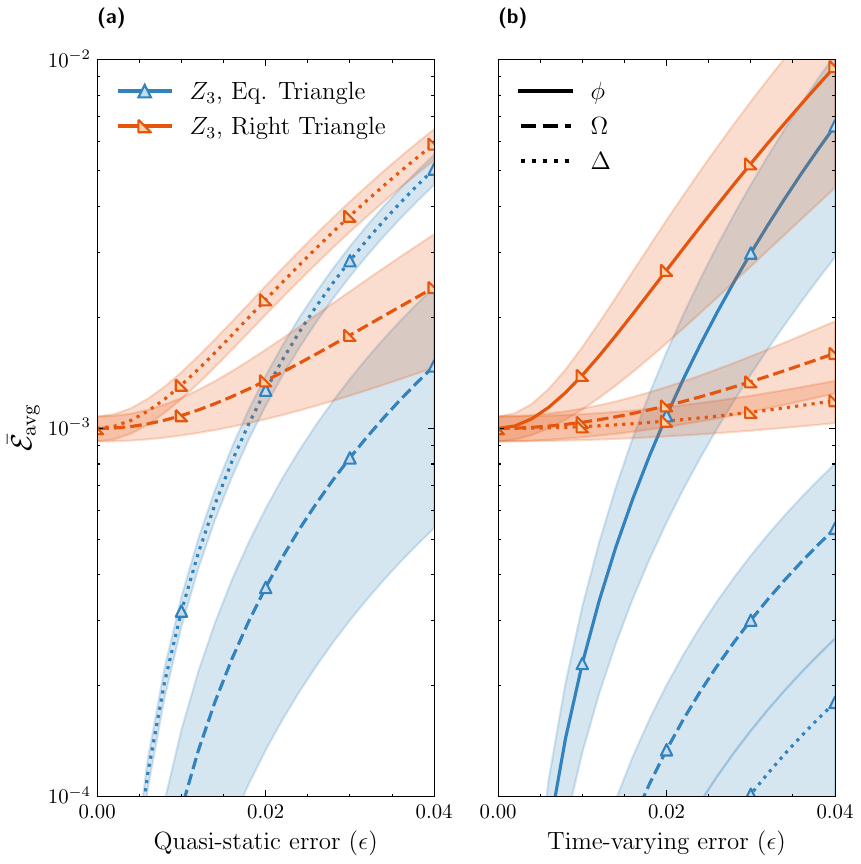}
  \caption{
Average gate infidelity under stochastic errors of control parameters for the $Z_3$ parity gate on the equilateral (blue) and right triangle (red) layouts.
(a) Quasi-static errors of the Rabi frequency $\Omega$ and detuning $\Delta$.
(b) Time-varying errors of the phase $\phi$, Rabi frequency $\Omega$ and detuning $\Delta$.
Each data point represents the average over 50 randomly sampled simulation runs, the shaded region indicates plus/minus one standard deviation. 
See main text for details.
}
\label{fig:robustness}
\end{figure}

We conclude by evaluating the robustness of the optimized control pulses under stochastic fluctuations in the laser field parameters, namely the phase, intensity, and frequency, which correspond to variations in \(\{\phi(t), \Omega(t), \Delta(t)\}\). These fluctuations are modeled as random draws from the Gaussian distribution \(\mathcal{N}(0, \sigma^2)\), with standard deviations \(\sigma = \epsilon \Omega_0\) for Rabi frequency \(\Omega\) and detuning \(\Delta\), and \(\sigma = \epsilon \pi\) for phase \(\phi\), where \(\epsilon\) denotes the relative error strength. We distinguish between quasi-static errors, where a single random perturbation is applied to an entire pulse (shot-to-shot noise), and time-varying errors, where a new random value is drawn for each time step $t_j$.

As a case study, we examine the \(\mathrm{Z}_3\) gate implemented on both the equilateral and right triangle configurations. 
We here base our analysis on the average gate fidelity $\mathcal{F}_{\rm avg}$
[see Eq.~\eqref{eq:F_avg} in App.~\ref{app:fidelity}] to not overemphasize the contributions of leakage out of the computational subspace~\cite{Pedersen2007, Fromonteil2023}.
In Fig.~\ref{fig:robustness}(a), we present the performance of the average gate infidelity $\mathcal{E}_{\rm avg} = 1 - \mathcal{F}_{\rm avg}$ under quasi-static errors in \(\Omega\) and \(\Delta\), for error strengths up to \(\epsilon = 0.04\).
The results indicate that the pulses are notably robust to Rabi frequency fluctuations, whereas detuning fluctuations lead to a moderate decrease of fidelity. 
This observation is consistent with earlier findings~\cite{Jandura2023}, which show that intrinsic detuning-robustness can only be achieved for vanishing Rydberg time $T_R$.
While we minimized $T_R$ by including it into the noise-aware cost function, Eq.~\eqref{eq:noise_aware_cost}, a non-zero $T_R$ is required for a successful implementation of any entangling gate. 

We omit static phase errors from this analysis, as global phase shifts do not affect the resulting gate unitary.
However, phase noise becomes significant when we consider time-varying errors, as illustrated in Fig.~\ref{fig:robustness}(b). In this case, the pulses maintain their robustness against Rabi frequency and detuning fluctuations, while phase noise emerges as the dominant source of infidelity. A more comprehensive investigation into the robustness properties of parity phase gates is left for future work.

\section{Conclusions}\label{sec:conclusion}
We have presented a detailed study based on optimal control techniques to realize multi-qubit parity phase gates natively in neutral atom QPUs through global phase modulation of the Rydberg excitation laser.
We have considered parity gates with up to four qubits and analyzed different spatial layouts of atom configurations.
We have shown that the quantum speed limit, indicating the minimal possible time for a successful gate operation, depends not only on the number of atoms involved, but also on the spatial configuration, in particular whether all atoms are equidistant, or not. 
In order to obtain pulse sequences that can be executed on real hardware with limiting control bandwidth we have additionally analyzed the effects of regularizing the pulse smoothness through the optimization cost function 
and found that high-fidelity gates can be realized with smooth pulses at the cost of increased pulse duration and reduced (but still high) fidelity. 

To mitigate fidelity loss from intrinsic noise sources, we enhanced our optimal control framework by incorporating noise-aware regularization terms into the cost function. This approach effectively minimizes the time that atoms spend in Rydberg states during control pulses.
By that, we could significantly reduce the sensitivity to noise sources such as spontaneous decay, photon recoil, and interaction-induced motional effects.
We demonstrated that this noise-aware strategy enhances gate fidelity without compromising experimental feasibility, especially in non-equidistant atomic configurations where trade-offs between infidelity contributions are more pronounced. Our results show that the native implementation of multi-qubit parity gates not only achieves superior fidelity compared to circuit decompositions, but also exhibits advantageous Pauli error biases beneficial for fault-tolerant quantum computing. 

Our findings highlight the potential of native, noise-aware gate design in advancing error-resilient quantum processors based on Rydberg platforms. Extending the optimal control framework to also include laser noise using response theory could further improve gate design for experimental realization \cite{Tao2024}.
Another interesting direction involves developing robust pulse protocols specifically tailored for multi-qubit phase gates, drawing upon well-established theoretical techniques \cite{Jandura2023, Fromonteil2023}.

\begin{acknowledgments}
We acknowledge useful discussions with Clemens Dlaska, Pedro Ildefonso, Alice Pagano, and Sebastian Weber. We specially thank Andrea Alberti, Andrew Byun, and Johannes Zeiher for fruitful discussions on experimental realization. 
This research is funded by the German Federal Ministry of Research, Technology and Space (BMFTR) within the project MUNIQC-ATOMS (Project No. 13N16080). This study was supported by the Austrian Research Promotion Agency (FFG Project No. FO999909249, FFG Basisprogramm) and by NextGenerationEU via FFG and Quantum Austria (FFG Project No.~FO999896208). 
This research was funded in part by the Austrian Science Fund (FWF) under Grant-DOI 10.55776/F71.
This project was funded within the QuantERA II Programme that has received funding from the European Union’s Horizon 2020 research and innovation programme under Grant Agreement No. 101017733. This publication has received funding under Horizon Europe Programme HORIZON-CL4-2022-QUANTUM-02-SGA via the project 101113690 (PASQuanS2.1). For the purpose of open access, the author has applied a CC BY public copyright license to any Author Accepted Manuscript version arising from this submission.
\end{acknowledgments}

\clearpage
\appendix
\section{Fidelity Metrics}\label{app:fidelity}

In this section, we present several key figures of merit (FOMs) used to evaluate how accurately a quantum operation $U$ implements a target gate $U_{\mathrm{target}}$. For clarity, we assume throughout that $U$ is unitary. The most commonly used performance metric is fidelity, which has multiple definitions depending on the context. We begin with the average gate fidelity, which quantifies the probability that the actual gate operation yields the same outcome as the target gate, averaged over all input states:
\begin{equation}
\begin{split}
    \mathcal{F}_{\mathrm{avg}} & = \int d\psi \left| \braketp{\psi}{O} \right|^2\\
    & = \frac{\big|\mathrm{Tr}\left[O\right]\big|^2 + \mathrm{Tr}\left[O O^\dagger\right]}{d(d+1)},
\end{split}
\end{equation}
where $d\psi$ denotes the normalized Haar measure over pure states, $d$ is the Hilbert space dimension, and $O = U_{\mathrm{target}}^\dagger U$. In practical situations, $U$ may act on a larger Hilbert space than the target gate, for instance when auxiliary levels are involved. In such cases, the fidelity should be computed in the relevant subspace using the substitution $O \rightarrow U_{\mathrm{target}}^\dagger P U P$, where $P$ is the projector onto the computational subspace of dimension $d$.

For phase gates, the average fidelity simplifies to:
\begin{equation}
    \begin{split}
    \mathcal{F}_{\mathrm{avg}} =& \frac{1}{2^{N}(2^N+1)} \Bigg( \left|\sum_{\mu\in\{0, 1\}^N}e^{-i\theta_{\mu}} \braketp{\mu}{U(T)}\right|^2 \\&+ \sum_{\mu\in\{0, 1\}^N}\left|e^{-i\theta_{\mu}} \braketp{\mu}{U(T)}\right|^2\Bigg)
    \end{split}
    \label{eq:F_avg}
\end{equation}

Another useful FOM, especially relevant in the context of Rydberg phase gates, is the generalized Bell state fidelity. This quantity measures the overlap between the states $U \ket{++\dots+}$ and $U_{\mathrm{target}} \ket{++\dots+}$ and is directly related to the first term of the average gate fidelity:
\begin{equation}
    \mathcal{F}_{\mathrm{Bell}} = \frac{\big|\mathrm{Tr}\left[O\right]\big|^2}{d^2}.
\end{equation}
For phase gates, this becomes:
\begin{equation}
    \mathcal{F}_{\mathrm{Bell}} = \frac{1}{4^{N}} \left|\sum_{\mu\in\{0, 1\}^N}e^{-i\theta_{\mu}} \braketp{\mu}{U(T)}\right|^2.
\end{equation}
It is important to note that while both Bell state fidelity and average gate fidelity degrade in the presence of population leakage outside the computational subspace, the latter is more sensitive to such effects~\cite{Pedersen2007, Fromonteil2023}.

\section{Dissipation-Induced Errors}
In this section, we analyze how dissipation, represented through jump operators in the Lindblad master equation, contributes to gate infidelity. We first characterize the types of errors that appear using the process matrix formalism~\cite{Korotkov2013}, particularly focusing on Pauli errors under the Pauli-twirling approximation (PTA). We then quantify how these errors reduce gate fidelity.

\subsection{Pauli Error Channels}\label{app:pauli_errors}

To analyze the error mechanisms, we consider the evolution of a quantum state $\rho_i$ through an ideal unitary gate $U$, followed by an error process characterized by a process matrix $\chi^{\mathrm{err}}$. The output state $\rho_f$ is given by:
\begin{equation}
    \rho_f = \sum_{m,n} \chi_{mn}^{\mathrm{err}} E_m U \rho_i U^\dagger E_n^\dagger,
\end{equation}
where $\{E_n\}$ represents the Pauli operator basis on the computational subspace, and the matrix $\chi^{\mathrm{err}}$ captures the nature and strength of the errors introduced after the gate operation. The normalization condition $\sum_{m,n} \chi_{mn}^{\mathrm{err}} E_m E_n^\dagger = \mathbb{I}$ ensures trace preservation. 

We model the open system dynamics using the Lindblad master equation
\begin{equation}
    \dot{\rho}(t) = -\frac{i}{\hbar} [H(t), \rho(t)] 
    + \sum_k \gamma_k \mathcal{D}[L_k]\rho,
\end{equation}
where $H(t)$ is the system Hamiltonian, and each $\mathcal{D}[L_k]$ represents a dissipative channel associated with a jump operator $L_k$. The Lindblad dissipator is defined as
\begin{equation}
    \mathcal{D}[L]\rho = L\rho L^\dagger - \frac{1}{2} L^\dagger L \rho - \frac{1}{2} \rho L^\dagger L,
\end{equation}
and models incoherent processes such as decay or dephasing. In the limit of weak dissipation $\gamma_k T \ll 1$ (with $T$ denoting the total duration of the evolution), errors due to different dissipation channels are additive and uncorrelated \cite{Korotkov2013}. This simplification allows us to treat each dissipation source independently in a perturbative expansion. We therefore expand the error process matrix $\chi^{\mathrm{err}}$ in the dissipation strength as
\begin{equation}
    \chi^{\mathrm{err}} = \chi^{\mathrm{err}}_{\mathrm{I}} + \sum_k \gamma_k \int_0^T dt\, \delta \chi^{\mathrm{err}}(t, L_k) + \mathcal{O}(\gamma_k^2T^2),
\end{equation}
where $\chi^{\mathrm{err}}_{\mathrm{I}}$ represents the contribution from the identity process, and the first-order correction due to each $L_k$ is given by:
\begin{equation}\label{eq:delta_chi}
\begin{split}
    \delta\chi^{\mathrm{err}}_{mn}(t, L) =& \frac{1}{d^2}\mathrm{Tr}\left[ L(t)E_m^\dagger\right] \mathrm{Tr}\left[ L(t)^\dagger E_n\right] \\
    &- \frac{1}{2d}\mathrm{Tr}\left[ L^\dagger(t) L(t) E_m^\dagger\right]\delta_{n0} \\
    &- \frac{1}{2d}\mathrm{Tr}\left[ L^\dagger(t) L(t) E_n\right]\delta_{m0},
\end{split}
\end{equation}
where $L(t) = U^\dagger(t) L U(t)$ denotes the jump operator in the interaction picture.

The first term in Eq.~\eqref{eq:delta_chi}, which affects the diagonal elements of the error process matrix, represents actual quantum jump events. In contrast, the second and third terms, which only influence the first row and column, stem from evolution without jumps. As a result, only the first term determines the probability $\chi_{nn}^{\mathrm{err}}$ of a specific Pauli error $E_n$ occurring. By averaging the noise channel over all Pauli operators, a method known as the Pauli twirling approximation~\cite{Zyczkowski2005, Geller2013}, the off-diagonal elements are eliminated, effectively reducing the noise model to a Pauli channel. This simplification enables efficient classical simulation of quantum systems under noise, which is particularly beneficial for quantum error correction.

The process fidelity $F_\chi$ can be read-off directly from the element of the (error) process matrix that corresponds to the Pauli errors being the identity, typically $F_{\chi} = \chi_{00}^{\mathrm{err}}$. Extracting this element from Eq.~\eqref{eq:delta_chi} defines how 
the total process infidelity $1 - F_{\chi}$ receives contributions from the underlying dissipation, i.e.
\begin{equation}
\begin{split}
    \delta \mathcal{E}_{\chi}(t, L) =
    &\frac{1}{d} \mathrm{Tr}\left[ L^\dagger(t) L(t) \mathbb{I}_{\mathrm{cmp}} \right] \\
    &- \frac{1}{d^2} \mathrm{Tr}\left[ L^\dagger(t) \mathbb{I}_{\mathrm{cmp}} \right] \mathrm{Tr}\left[ L(t) \mathbb{I}_{\mathrm{cmp}} \right],
\end{split}
\end{equation}
where $\mathbb{I}_{\mathrm{cmp}}$ is the projector onto the computational subspace.

\subsection{Fidelity Reduction}
The average gate fidelity $\mathcal{F}_{\mathrm{avg}}$ quantifies the overall performance of a quantum gate under noise. In the presence of weak dissipation, the loss in fidelity can be approximated by integrating the instantaneous fidelity reduction over time
\begin{equation}
    1 -\mathcal{F}_{\rm avg} \coloneq \mathcal{E}_{\mathrm{avg}} = \sum_k \gamma_k \int_0^T dt\, \delta \mathcal{E}_{\mathrm{avg}}(t, L_k) + \mathcal{O}(\gamma_k^2T^2),
\end{equation}
with the integrand given by
\begin{equation}\label{eq:E_avg}
\begin{split}
    \delta \mathcal{E}_{\mathrm{avg}}(t, L) = 
    &\frac{1}{d} \mathrm{Tr}\left[ L^\dagger(t) L(t) \mathbb{I}_{\mathrm{cmp}} \right] \\
    &- \frac{1}{d(d+1)} \mathrm{Tr}\left[ L^\dagger(t) \mathbb{I}_{\mathrm{cmp}} L(t) \mathbb{I}_{\mathrm{cmp}} \right] \\
    &- \frac{1}{d(d+1)} \mathrm{Tr}\left[ L^\dagger(t) \mathbb{I}_{\mathrm{cmp}} \right] \mathrm{Tr}\left[ L(t) \mathbb{I}_{\mathrm{cmp}} \right].
\end{split}
\end{equation}
A detailed derivation can be found in Ref.~\cite{Abad2025}.

As demonstrated in Ref.~\cite{Wood2018}, the average gate infidelity $\mathcal{E}_{\mathrm{avg}}$ is directly related to the process infidelity $\mathcal{E}_{\chi}$ via:
\begin{equation}
    \mathcal{E}_{\mathrm{avg}} = \frac{d \mathcal{E}_{\chi} + \mathcal{L}}{d + 1},
\end{equation}
where $\mathcal{L}$ represents the leakage rate, i.e., the probability that population escapes the computational subspace:
\begin{equation}
    \mathcal{L} = \sum_k \gamma_k \int_0^T dt\, \delta \mathcal{L}(t, L_k) + \mathcal{O}(\gamma_k^2T^2),
\end{equation}
with
\begin{equation}
\begin{split}
    \delta \mathcal{L}(t, L) = 
    &\frac{1}{d} \mathrm{Tr}\left[ L^\dagger(t) L(t) \mathbb{I}_{\mathrm{cmp}} \right] \\
    &- \frac{1}{d} \mathrm{Tr}\left[ L^\dagger(t) \mathbb{I}_{\mathrm{cmp}} L(t) \mathbb{I}_{\mathrm{cmp}} \right].
\end{split}
\end{equation}

As a specific example of dissipation, consider the Rydberg decay, described by the jump operator $L_k = \ketbrap{1_k}{r_k}$. In this case, the dominant contribution to the average gate infidelity turns out to be proportional to the total time spent in the Rydberg state, denoted $T_R$, given by
\begin{equation}
    T_R = \frac{1}{d} \sum_{k=1}^N \int_0^T dt\, \mathrm{Tr}\left[ L_k^\dagger(t) L_k(t) \mathbb{I}_{\mathrm{cmp}} \right], \, \text{for } L_k = \ketbrap{1_k}{r_k}.
\end{equation}

Other contributions to the infidelity are negligible in this scenario. To illustrate, consider the evolution of a system initialized in a computational basis state $\ket{\psi_\mu}$. Due to coupling between the Rydberg and $\ket{1}$ states, the system's state evolves as $\ket{\psi(t)} = c_\mu \ket{\psi_\mu} + c_r \ket{\psi_r}$, where $\ket{\psi_r}$ lives in the Rydberg subspace associated with $\mu$. Assuming a fully blockaded regime, $\ket{\psi_r}$ becomes a superposition of singly-excited Rydberg states. Neglecting leakage to non-computational states, the normalization condition $|c_\mu|^2 + |c_r|^2 = 1$ holds.
Analyzing the terms in Eq.~\eqref{eq:E_avg} under Rydberg decay, the first term scales with $\frac{1}{d}|c_r|^2$, while the second and third scale as $\frac{1}{d(d+1)}|c_r c_\mu|^2 = \frac{1}{d(d+1)}|c_r|^2(1 - |c_r|^2)$. Therefore, the average gate infidelity is approximately:
\begin{equation}
    \mathcal{E}_{\mathrm{avg}}(L_k = \ketbrap{1_k}{r_k}) \approx \gamma_d T_R - \frac{2}{d+1}\gamma_d T_R + \mathcal{O}(\gamma_d^2T^2),
\end{equation}
where the leading term comes from no-jump evolution. The subleading term arises from jump processes and becomes negligible as the Hilbert space dimension $d$ increases. Thus, to first order in $\gamma_d T$, the infidelity from decay is bounded by:
\begin{equation}
    \mathcal{E}_{\mathrm{avg}}(\ketbrap{1_k}{r_k}) \leq \gamma_d T_R.
\end{equation}
This approximation has been validated in prior work~\cite{Robicheaux2021, Jandura2022, Pagano2022, Poole2025}, and justifies the description of the Rydberg decay in terms of a complex term in the Hamiltonian, as we have done in the main text [c.f. Eq.~\eqref{eq:spin-phonon_ham}].

\section{Enhancing Gate Performance: Role of Rabi Frequency}\label{app:rabi}
\begin{figure}[!t]
  \centering
  {\includegraphics[width=0.8\linewidth]{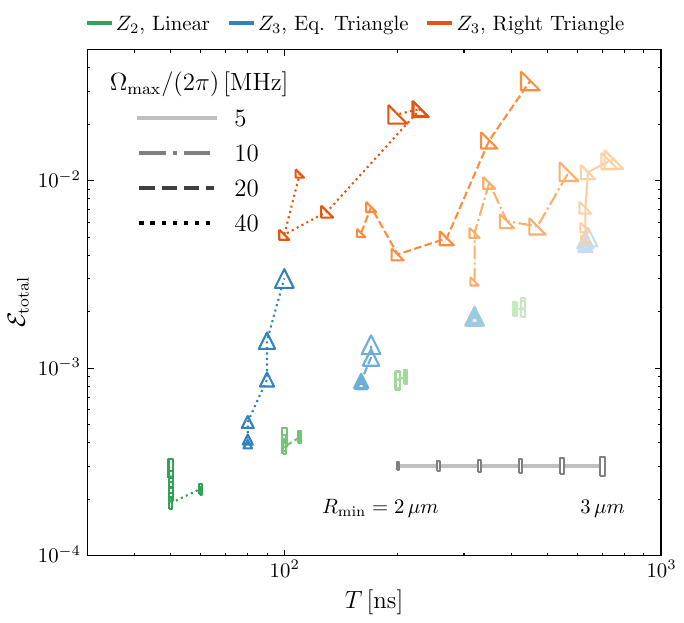}}
  \caption
    {%
      Total gate infidelity as a function of gate duration for different maximum Rabi frequencies \(\Omega_{\max}/(2\pi) = 5,\, 10,\, 20,\, 40\, \mathrm{MHz}\), indicated by distinct lines and color gradients. Marker shapes and colors correspond to different atomic configurations, while marker sizes reflect interatomic spacings of \(2.0,\, 2.2,\, 2.4,\, 2.6,\, 2.8,\, 3.0\, \mathrm{\mu m}\). Each data point represents the lowest infidelity obtained from 20 optimization runs using the noise-aware cost function, Eq.~\eqref{eq:noise_aware_cost}, and reflect the best performance under each set of conditions.
    }%
    \label{fig:varying_rabi}
\end{figure}

In this section, we provide a perspective on how the performance of multi-qubit parity phase gates might be further enhanced to support deeper quantum circuit computations. One approach involves increasing the Rabi frequency, which can reduce gate durations. However, the resulting impact on overall gate fidelity is not straightforward. For example, considering errors from Rydberg decay described by 
\(
\mathcal{E}_{\mathrm{decay}}=\frac{\gamma_d(\Omega_{\max} T_R)}{\Omega_{\max}},
\)
increasing \(\Omega_{\max}\) can lower the infidelity if the numerator remains constant. 
This could be achieved by scaling the pulse's duration $T$ indirectly proportional to $\Omega_{\rm max}$, under the assumption that the ratio of the Rydberg time to the pulse duration remains constant, i.e. $T_R/T = \rm{const.}$.
The latter condition, however, can only be maintained when the interaction strength between the atoms can also be increased proportional to $\Omega_{\rm max}$. 
Otherwise increased finite-blockade effects can introduce additional errors, which scale with $\Omega_{\rm max}^2$, even in the absence of noise, as discussed in Ref.~\cite{Jandura2022}.
Increasing the interaction strength, without modifying other relevant quantities (e.g. the decay rate) is possible by reducing the inter-atomic distances.
With current technology, however, minimal distances below a few micrometers are not achievable, eventually leading to a trade-off between increasing $\Omega_{\max}$ and introducing additional errors from increased finite-blockade effects.
Similar considerations hold for errors introduced from photon recoil, which scales $\propto T_R^2$, and from the VdW force, $\propto T_{RR}^2$. 

To investigate this trade-off in more detail, we conducted numerical simulations for 2- and 3-qubit gates, as shown in Fig.~\ref{fig:varying_rabi}. The simulations explore a range of maximum Rabi frequencies up to \(\Omega_{\max}/(2\pi)= 40\,\mathrm{MHz}\), with interatomic distances spanning from \(2\) to \(3\,\mathrm{\mu m}\). The results indicate that increasing the Rabi frequency generally leads to shorter gate durations. However, the overall infidelity \(\mathcal{E}_{\mathrm{total}}\) exhibits more complex behavior. In particular, for larger interatomic separations, errors associated with finite blockade strength become increasingly significant. This effect is especially pronounced in non-equidistant geometries, such as the right triangle, where the increase in both \(T_R\) and \(T_{RR}\) influences all error contributions. In summary, these findings point to a viable route toward achieving fidelities beyond the \(0.999\) threshold for multi-qubit Rydberg gates, especially when leveraging noise-aware optimal control strategies.

\section{Parametrized parity gates}\label{app:different_angles}
\begin{figure}[!t]
  \centering
    {\includegraphics[width=0.9\linewidth]{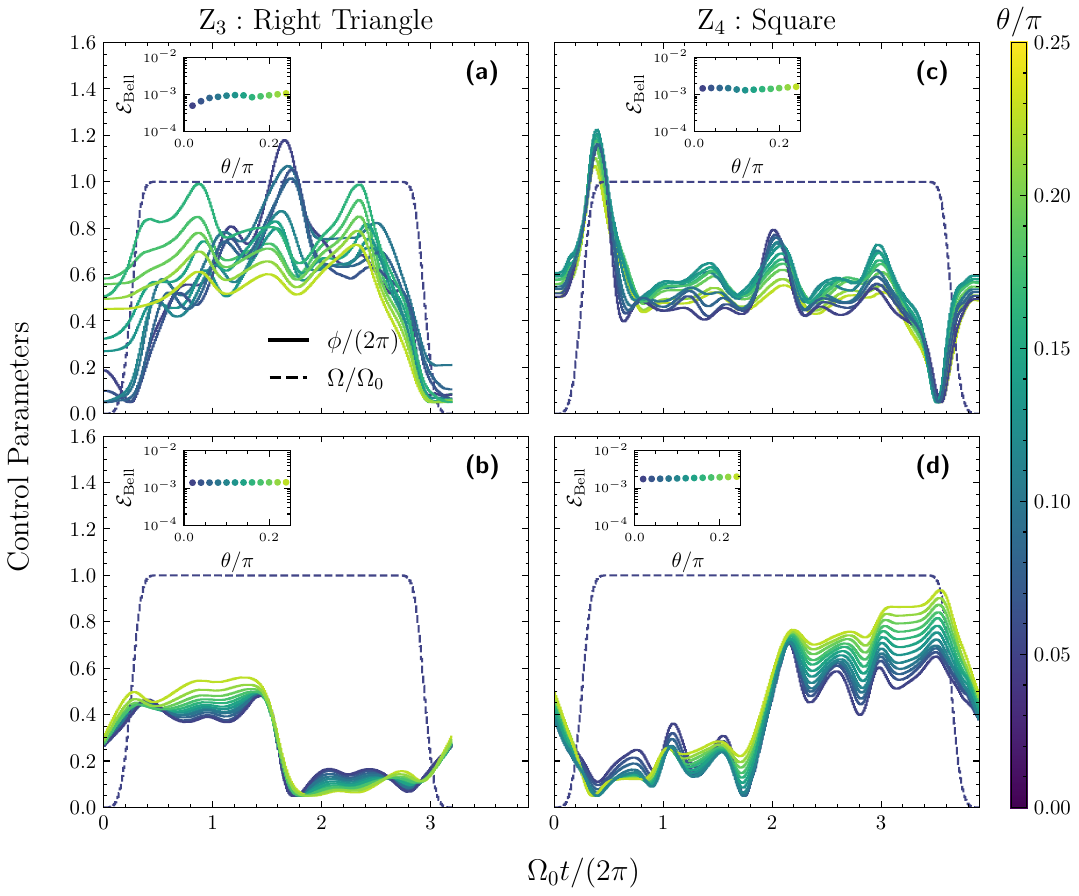}}
  \caption
    {%
    Optimized pulse profiles for multi-qubit parity gates with varying angles \(\theta \in [0, \pi/4]\), obtained by iterative oprtimization from a pre-computed optimal pulse at \(\theta = \pi/4\). Figures (a) and (b) show two distinct families of pulses for the \(\mathrm{Z}_3(\theta)\) gate on a right triangle atomic configuration, where anti-symmetric pulses of the laser phase are enforced in (b) during the optimization.
    Figures (c) and (d) display corresponding results for the \(\mathrm{Z}_4(\theta)\) gate in a square configuration. The insets present the corresponding Bell-state infidelities, indicating that high fidelity is maintained across the range of \(\theta\).
    For both gates, the anti-symmetric pulses result in a more structured pulse evolution with respect to the target angle $\theta$.
    }%
    \label{fig:different_angles}
\end{figure}

In the main text, we have focused on maximally-entangling parity phase gates $\mathrm{Z}_N(\theta=\frac{\pi}{4})$. 
Here, we briefly discuss the construction of optimal parity phase gates for different values of the angle \(\theta\). A straightforward method to generalize the \(\mathrm{Z}_N(\theta=\frac{\pi}{4})\) gate to arbitrary angles involves using its already optimized parameters as the initial guess for a nearby angle, \(\theta^\prime = \frac{\pi}{4} + \delta\theta\). By applying this procedure iteratively, one can obtain optimized pulses for a continuous range of angles, promoting smooth transitions between them. Using this approach, we present numerical results in Fig.~\ref{fig:different_angles}, illustrating pulse profiles for multi-qubit parity gates across different \(\theta\) values. We particularly focus on non-equidistant atomic configurations, which typically result in more intricate pulse structures. The outcomes show that the pulse shapes adjust rather smoothly as \(\theta\) varies from the maximally entangling case \(\frac{\pi}{4}\) down to \(0\), while maintaining fidelity close to the original value for a fixed gate duration [see insets]. In the second row of the figure, we present a constrained pulse design where the laser phase is required to be anti-symmetric with respect to the pulse midpoint. This constraint yields very homogeneous transitions of the pulse profile, which can be particularly advantageous in scenarios where \(\theta\) is dynamically tuned.

Other strategies for parameterizing quantum gates are also worth noting. For example, one can precompute optimized pulses for a discrete set of \(\theta\) values, and interpolate between pulses within the same family to obtain control parameters for arbitrary angles, as proposed in Ref.~\cite{DeKeijzer2024}. Additionally, piecewise-constant pulses derived via optimal control can, in principle, be transformed into analytic functions with fewer free parameters, which may facilitate experimental implementation \cite{Jandura2022, Ma2023}. Finally, we note that the experimental characterization of such parameterized parity gates could be approached using non-Clifford learning techniques, as suggested in Ref.~\cite{Layden2024}. We leave the exploration of this direction for future work.


\clearpage
\bibliographystyle{apsrev4-2}
\bibliography{bibliography.bib}
\clearpage

\end{document}